\documentclass[conference]{IEEEtran}
\IEEEoverridecommandlockouts
\usepackage{cite}
\usepackage{subfig}
\usepackage{enumitem}
\usepackage{multirow}
\usepackage{amsmath,amssymb,amsfonts}
\usepackage{algorithmic}
\newtheorem{example}{Example}
\newtheorem{definition}{Definition}
\newtheorem{lemma}{Lemma}
\newtheorem{problem}{Problem}
\usepackage[linesnumbered,ruled]{algorithm2e}
\newcommand{\ssymbol}[1]{^{\@fnsymbol{#1}}}
\newenvironment{proc}[1][htb]
  {
   \begin{algorithm}%
  }{\end{algorithm}}

\usepackage{graphicx}
\usepackage{textcomp}
\usepackage{xcolor}
\def\BibTeX{{\rm B\kern-.05em{\sc i\kern-.025em b}\kern-.08em
    T\kern-.1667em\lower.7ex\hbox{E}\kern-.125emX}}
\begin{document}

\title{HushRelay: A Privacy-Preserving, Efficient, and Scalable Routing Algorithm for Off-Chain Payments}

\author{
	Subhra Mazumdar$\ast$, Sushmita Ruj$\ddag$, Ram Govind Singh$\dag$, Arindam Pal$\ddag\bigstar$\\
	$\ast$ Indian Statistical Institute Kolkata, India, Email: subhram\_r@isical.ac.in \\
	$\bigstar$Cyber Security CRC, Sydney, New South Wales, Australia\\
	$\ddag$ CSIRO, Data61, Australia, Email: \{sushmita.ruj, arindam.pal\}@data61.csiro.au\\
	$\dag$ ICERT, Ministry of Electronics and Information Technology, India, Email: ramgovind.2010@gmail.com
}
\maketitle

\begin{abstract}
Payment channel networks (PCN) are used in cryptocurrencies to enhance the performance and scalability of off-chain transactions. Except for opening and closing of a payment channel, no other transaction requests accepted by a PCN are recorded in the Blockchain. Only the parties which have opened the channel will know the exact amount of fund left at a given instant. In real scenarios, there might not exist a single path which can enable transfer of high value payments. For such cases, splitting up the transaction value across multiple paths is a better approach. While there exists several approaches which route transactions via several paths, such techniques are quite inefficient, as the decision on the number of splits must be taken at the initial phase of the routing algorithm (e.g., SpeedyMurmur \cite{speedymurmur}).

Algorithms which do not consider the residual capacity of each channel in the network are susceptible to failure. Other approaches leak sensitive information, and are quite computationally expensive \cite{silentwhispers}. To the best of our knowledge, our proposed scheme \textit{HushRelay} is an efficient privacy preserving routing algorithm, taking into account the funds left in each channel, while splitting the transaction value across several paths. Comparing the performance of our algorithm with existing routing schemes on real instances (e.g., Ripple Network), we observed that \emph{HushRelay} attains a success ratio of 1, with an execution time of 2.4 sec. However, \emph{SpeedyMurmur} \cite{speedymurmur} attains a success ratio of 0.98 and takes 4.74 sec when the number of landmarks is 6. On testing our proposed routing algorithm on the Lightning Network, a success ratio of 0.99 is observed, having an execution time of 0.15 sec, which is 12 times smaller than the time taken by \emph{SpeedyMurmur}.

\end{abstract}



%

\section{Introduction}
Cryptocurrencies like Bitcoin \cite{nakamoto2008bitcoin} have gained popularity as an alternative method of payment. Blockchain, a cryptographically secure, tamper proof ledger, forms the backbone of such decentralized network, guaranteeing pseudonymity of participant. The records stored in this distributed ledger can be verified by anyone in the network. Consensus algorithms like Proof-of-Work \cite{nakamoto2008bitcoin}, \cite{o2014bitcoin}, \cite{bano2017consensus}, Proof-of-Stake \cite{king2012ppcoin}, \cite{li2017securing}) are used for reaching agreement on state change in the ledger across the network participants. However,  computation time taken by such consensus algorithm is the major bottleneck in scalability of blockchain based transaction \cite{croman2016scaling}, \cite{poon2016bitcoin}. To be at par with traditional methods of payment like Visa, PayPal, scaling blockchain transactions is an important concern which needs to be addressed, without compromising on the privacy.

Several solutions like sharding \cite{luu2016secure}, \cite{gencer2016service}, alternate consensus architecture \cite{kiayias2017ouroboros}, \cite{miller2014permacoin}, \cite{buterin2017casper}, \cite{park2018spacemint}, \cite{eyal2016bitcoin}, \cite{pass2017hybrid}, side-chains \cite{back2014enabling} have been proposed in Layer-one. But this requires revamping the trust assumptions of the base layer and changing the codebase. A more modular approach is exploring scalability in Layer-two \cite{gudgeon2019sok}. It massively cuts down data processing on the blockchain by running computations off-chain. The amount of data storage on Layer-one is minimized. Taking transactions off the base layer, while still anchored to it, would free up processing resources to do other things. Also Layer-two relies on Layer-one for security. Several solutions like \cite{decker2018eltoo}, \cite{decker2015fast},  \cite{luu2016secure} have been proposed. \textit{Payment Channel} \cite{decker2015fast} stands out as a practically deployable answer to the scalability issue. 

Any two users, with mutual consent, can open a payment channel by locking their funds in a deposit. Users can perform several off-chain payments between each other without recording the same on blockchain. This is done by locally agreeing on the new deposit balance, enforced cryptographically by smart contracts \cite{poon2016bitcoin}, key based locking \cite{malavoltamulti} etc. Whenever one of the party wants to close the payment channel, it broadcasts the transaction on blockchain with the final balance. None of the parties can afford to cheat by claiming payment for an older transaction. Opening of new payment channel between parties which are not connected directly has its overhead where funds get locked for a substantial amount of time. This can be avoided by leveraging on the set of existing payment channels for executing a transaction, proving beneficial in terms of resource utilization. These set of payment channels form the \textit{Payment Channel Network (PCN)}. Several problem such as routing, security and interoperability needs to be addressed in such a network.


The major challenge in designing any protocol for PCN is to ensure privacy of payer and payee and hiding the payment value transferred. No party, other than the payer and payee, should get any information about the transaction. Thus any routing algorithm designed for such a network must be decentralized, where individual nodes take decision based on the information received from its neighbourhood. Several distributed routing algorithms exists but they suffer from various disadvantages - Elias et al. \cite{elias} requires a single node to maintain list of active vertices for executing push relabel algorithm on single source-sink pair, Flare  \cite{prihodko2016flare} requires intermediate users to reveal the current capacity of their payment channels to the sender for computation of the maximum possible value to be routed through a payment path, Canal  \cite{viswanath2012canal} entrusts a single node for computing maximum flow in a graph. Landmark-based routing algorithms \cite{silentwhispers}, \cite{speedymurmur} decide the number of landmarks by trial and error. If the total number of landmarks is $k$, then the payment value is split into $k$ microtransactions randomly without considering the nature of the graph. Such a myopic approach for routing each microtransaction may result in failure as it does not allow optimal utilization of the available capacities present across multiple paths.

It was first mentioned in Elias et al. \cite{elias} that push relabel fits better as a routing algorithm for PCN  because it proceeds locally, taking into account the residual capacity of each payment channel. However, the push relabel algorithm used for single source-sink pair \cite{elias} is not decentralized in nature. A distributed version of the same was implemented in their paper for multiple source-sink pair but it is not well defined. It is not clear how many payment transfer can be allowed at a time through a channel. Further, it was assumed that each payment value for a source-sink pair is unsplittable. This assumption does not work in real life since the payment value might be higher than the bottleneck capacity of a single path. Deciding feasible routes even for a single payment transfer is an involved process in a distributed network. This motivated us to design a new routing algorithm for PCN which is privacy-preserving, efficient as well as scalable.

\vspace{-0.2cm}
\subsection{\textbf{Our Contributions}}
The following contributions have been made in this paper :
\begin{itemize}[leftmargin=*]

\vspace{-0.1cm}
\item We have proposed a privacy preserving distributed routing algorithm, \textit{HushRelay}, in payment channel network. 

\item We have implemented the scheme and its performance has been compared with SpeedyMurmur \cite{speedymurmur} in terms of \textit{success ratio} and \textit{time taken to route (TTR)} a payment. Testing was done on real instances of Ripple Network and Lightning Network \footnote{In the absence of widespread PCN, we use the statistics of such real instances to create the network} and it is observed that \emph{HushRelay} attains a success ratio of 1 in both the cases. However \emph{SpeedyMurmur} attained a maximum success ratio of 0.9815 and 0.907 respectively, when number of landmarks is 6. The time taken to execute the routing algorithm in Ripple like Network and Lightning Network are 2.4s and 0.15189s for \emph{HushRelay} but it takes 4.736s and 1.937s for \emph{SpeedyMurmur}. These statistics justify our claim of the algorithm being efficient and scalable. The code is given in \cite{Code}.

\item The proposed routing algorithm is modular and it can be combined with any other privacy preserving payment protocol.
\end{itemize}
 
\subsection{\textbf{Organization}}
Section \ref{2} discusses the state-of-the-art in PCN. Section \ref{3} gives a brief overview of the preliminaries. Section \ref{4} defines the problem statement and Section \ref{5} provides discusses \emph{HushRelay} with Section \ref{route} dealing with Generic Construction and Section \ref{correct} providing the proof of correctness. Performance analysis of each subprotocol of \textit{HushRelay} has been stated in Section \ref{6} and Section \ref{7} concludes the paper.
\vspace{-0.2cm}
\section{Related Work}
\label{2}
A payment channel network is a peer-to-peer, path based transaction (PBT) network, where each party operates independent of other parties. Several P2P path-based transaction networks such as such as the Lightning Network for Bitcoin \cite{poon2016bitcoin}, the Raiden Network for Ethereum \cite{raiden}, SilentWhispers \cite{silentwhispers}, InterLedger \cite{thomas2015protocol}, Atomic-swap \cite{atomic}, TeeChain \cite{lind2017teechain} etc. have been developed over the years. Perun \cite{dziembowski2017perun} proposes a more efficient network structure which is built around payment hubs. An extension of payment channel, State Channel Network \cite{dziembowski2018general}, not only supports off chain payment but allows execution of complex smart contract. Spider network \cite{sivaraman2018routing} adheres to a packet-switched architecture for payment channel network. Payment is split into several transaction units and it is transmitted over time across different paths. However the split does not take into account the bottleneck capacity of each path which might lead to failure of payment. BlAnC \cite{panwar2019blanc}, a fully decentralized blockchain-based network, has been proposed which transfers credit between a sender and receiver on demand. 

Till date, the routing algorithms proposed for payment channel network are as follows : Canal \cite{viswanath2012canal} - uses a centralized server for computing the path, Flare \cite{prihodko2016flare} - requires intermediate nodes to inform source node about their residual capacity, SilentWhispers \cite{silentwhispers} - a distributed PBT network without using any public ledger, SpeedyMurmur \cite{speedymurmur} - a privacy preserving embedded based routing, extending Voute \cite{roos2016voute}, depending on presence of landmark nodes. SpeedyMurmur is the most relevant privacy preserving distributed routing algorithm. However, it makes use of repeated trials to figure out a suitable split of the total transaction value across multiple paths. Elias et al. \cite{elias} proposed an extended push relabel for finding payment flow in the payment network. They are the first to point out the flaw in assumption of considering transaction \textit{unsplittable} for existing routing techniques. In real life, splitting of fund across multiple path is inevitable since the bottleneck capacity of a single path may be lower than the total value of fund transfer. Later, a distributed approach for PCN routing, CoinExpress \cite{yu2018coinexpress}, was proposed for finding routes that fulfill payment with higher success ratio. A routing algorithm based on swarm intelligence, ant colony optimization \cite{grunspan2018ant} has been explored. Hoenisch et al. \cite{hoenisch2018aodv} proposed an adaptation of an Ad-hoc On-demand Distance Vector (AODV)-based routing algorithm which supports different cryptocurrencies allowing transactions across multiple blockchains. We observe that none of the past works provide an efficient and secure routing algorithm. It is either susceptible to leaking of sensitive information or there exist a central entity controlling the routing algorithm.

\section{Background}
\label{3}
In this section, we provide the required background on payment channel network. 
The terms source/payer means the sender node. Similarly, sink/payee/destination means the receiver node and transaction means payment transfer.

\subsection{\textbf{Payment Channel Network}}
\begin{definition}
\label{basic}
A Payment Channel Network (PCN) \cite{malavolta} is defined as a bidirected graph $G:=(V,E)$, where $V$ is the set of accounts dealing with cryptocurrency and $E$ is the set of payment channels opened between a pair of accounts. A PCN is defined with respect to a blockchain. Only opening and closing of payment channel gets recorded on blockchain apart from disputed transactions where settlement is done by broadcasting the transaction on blockchain.
\end{definition}

Basic operations of PCN \cite{malavolta}-  
\begin{itemize}[leftmargin=*]
\item \texttt{openPaymentChannel$(v_1,v_2,\alpha,t,m)$} : For a given pair of accounts  $v_1,v_2 \in V$, channel capacity $\alpha$ (initial balance escrowed), timeout value of $t$ and processing fee charged $m$, \texttt{openPaymentChannel} creates a new payment channel $(id_{(v_1,v_2)},\alpha,t,m) \in E$, where $id_{(v_1,v_2)}$ is the channel identifier, provided both $v_1$ and $v_2$ has authorized to do so and the funds contributed by each of them sum up to value $\alpha$. 

\item \texttt{closePaymentChannel$(id_{(v_1,v_2)},\tilde{\alpha})$} : Given a channel identifier $id_{(v_1,v_2)}$ with balance $\tilde{\alpha}$, \texttt{closePaymentChannel} removes the channel from $G$ provided it is authorized to do so by both $v_1,v_2 \in V$. The balance $\tilde{\alpha}$ gets written on blockchain and this amount is distributed between $v_1$ and $v_2$ as per the net balance recorded.

\item \texttt{payVal$(p(s,r),val)$} : $p(s,r)$ denotes a path between sender $s$ and receiver $r$. It is defined by a set of identifiers $id_{(s,v_1)},id_{(v_1,v_2)},\ldots,id_{(v_n,r)}$, $s,v_1,v_2,\ldots,v_n,r \in V$, having enough credit to allow transfer of $val$ from $s$ to $r$, if for each payment channel denoted by $id_{(v_i,v_{i+1})}$ has capacity of at least $\beta  \geq val_i', val_i'=val+\Sigma_{j=i+1}^{n} fee(v_j),   0 \leq i \leq n, v_0=s \ and \ v_{n+1}=r$, where $fee(v_j)$ is the processing fee charged by each intermediate node $v_j$ in $p(s,r)$. A successful \texttt{payVal} operation leads to a decrease of capacity of each payment channel $id_{(v_i,v_{i+1})}$ by $val_i'$. Else the capacity of the channel remains unaltered. 
\end{itemize}

\subsection{\textbf{Payment Flow problem}}
\label{flowpay}
Consider a directed graph $G:=(V,E) : \  n=|V|, m=|E|, m \geq n-1$, having two distinguished vertices, source $s \in V$, sink $r \in V, s \neq r$, as a \textit{flow network.} For a pair of vertices $v,w$, distance from $v$ to $w$ in graph $G$ is defined by $d_G(v,w)$, the minimum number of edges on the path from $v$ to $w$; if there is no path from $v$ to $w$, $d_G(v,w)=\infty$. A positive real-valued capacity $c(v,w)$, defined by $c : E \rightarrow \mathbb{R}$, is the amount of funds that can be
transferred between two nodes sharing an edge. For every edge $(v,w) \in E$ ; if $(v,w) \not \in E$, then $c(v,w)=0$. A flow $f$ on $G$ is a real-valued function  on vertex pairs satisfying the constraints \cite{tarjan}, \cite{thuy2005distributed} :
\begin{equation}
\label{eq1}
\begin{matrix}
f(v,w) \leq c(v,w), \ \forall (v,w) \in V \times V \ \textrm{(capacity)}, \\
f(v,w)=-f(w,v), \ \forall (v,w) \in V \times V \ \textrm{(antisymmetry)}, \\
\Sigma_{u \in V} f(u,v)=0 \ \forall v \in V-\{s,r\} \ \textrm{(flow-conservation)}, \\
\end{matrix}
\end{equation}
The net flow into the sink is given by $f$, where:
\begin{equation}
f=\Sigma_{v \in V} f(v,r)
\end{equation}
A payment channel network can be mapped to flow network with channels forming the edges and funds locked on each channel can be considered as the edge capacity. Finding the maximum flow value from source to sink for a flow network is termed as the \textit{Maximum Flow problem}. In the context of PCN, given a payment value $val$, one has to find a feasible flow from payer to payee, which is termed here as \textit{Payment Flow problem}. Any max-flow algorithm with subtle modifications can be applied here, taking into account the preflow $f$ of each vertex (except the source and sink) on the network. A preflow is a real-valued function on a vertex pair which satisfies the first two constraints of Eq. \ref{eq1} and a weaker form of the third constraint :
\begin{equation}
\Sigma_{u \in V} f(u,v)\geq 0, \ \forall v \in V-\{s,r\} \ \textrm{(non-negativity constraint)},
\end{equation}

A residual capacity of an edge $(v,w) \in E$ is the amount of capacity remaining after the preflow $f$, i.e. $c(v,w)-f(v,w)$ and it is denoted by $r_f(v,w)$. A residual graph $G_f=(V,E_f)$ for a preflow $f$ is the graph whose vertex set is $V$ and edge set $E_f$ is the set of residual edges $(v,w) \in E \ : \ r_f(v,w)>0$. 
The \textit{flow excess e(v)} of a vertex $v$ is the net balance of funds in node $v$ denoted by $\Sigma_{u \in V} f(u,v)$. The algorithm ends with all vertices except $s$ and $r$ having zero excess flow. If sink is unreachable or if the network does not have adequate capacity for transferring the amount $val$, then the excess value is pushed back to source $s$. 

\section{Problem Statement}
\label{4}
It is not always possible to route the transaction across a single path as the value may be quite high compared to minimum capacity of the designated path. Hence it is better to find set of paths such that the total amount to be transferred is split across each such path. We define the problem as follows -
\begin{problem}
\textit{Given a payment channel network $G(V,E)$, a transaction request ($s,r,val$) for a source-sink pair $(s,r)$, the objective is to find a set of paths $p_1,p_2,\ldots,p_m$ for transferring the fund from $s$ to $r$ such that $p_1$ transfers $val_1$, $p_2$ transfers $val_2,\ldots,$ $p_m$ transfers $val_m : val=\Sigma_{i=1}^m val_i$ without violating \textit{transaction level privacy} i.e. neither the sender nor the receiver of a particular transaction must be identified as well as hiding the actual transaction value from intermediate parties.}
\end{problem}

\section{Our Proposed Construction}
\label{5}
In this section we provide a detailed overview of the routing algorithm, \textit{HushRelay}. The payment network comprises set of payment channels denoted by channel identifier $id_{(i,j)}, (i,j)\in E$.
We describe state the model and the assumptions made. 

\subsection*{\textbf{Network Model and its Assumptions}}
\begin{itemize}[leftmargin=*]
    \item The network is static i.e. no opening of new payment channel or closing of existing payment channel is considered.
    \item The topology of the network is known by any node in the network since any opening or closing of channel is recorded on the blockchain.
    \item Atmost one timelock contract is allowed to be established on a payment channel at a time.
    \item Sender of a payment chooses set of paths to the receiver according to her own criteria.
    \item  The current value on each payment channel is not published but instead kept locally by the users sharing a payment channel. 
    \item Pairs of users sharing a payment channel communicate through secure and authenticated channels
(such as TLS).

\end{itemize}  

\subsection{Generic Algorithm}
\label{route}
Since we consider PCN as a flow network, for solving the \textit{payment flow problem} in the given network for executing a transaction request $(s,r,val)$, we propose a routing algorithm inspired from distributed push relabel algorithm stated in \cite{tarjan}, \cite{thuy2005distributed}. The algorithm proceeds locally by exchange of messages between neighbouring nodes. No single entity controls the flow in the network.

Before discussing the algorithm, we briefly describe the \textit{Push Relabel} algorithm for a single source-sink pair (as stated in \cite{tarjan}):
\begin{itemize}[leftmargin=*]
\item The instruction \textit{push} redirects the excess flow of a vertex to the sink via its neighbouring vertices. The amount of excess flow that can be \textit{pushed} from a vertex $v$ to one of its neighbouring vertex $w$ is $\delta=min(e(v),r_{f(v,w)})$, where $r_{f(v,w)}$ is the residual capacity of edge $(v,w)$. The value $\delta$ is added to the preflow value $f(v,w)$ (subtracted from $f(w,v)$) and subtracted from $e(v)$. Any push which results in zero residual capacity of the edge is said to be \textit{saturating}. 

\item A valid labeling function $d : V \rightarrow I^{+}\cup\{0,\infty\}$ is used for estimating the distance of a vertex $v$ from sink $r$. $d(s)=n, d(r)=0$ and $d(v)\leq d(w)+1$ for every residual edge $(v,w)$. The label $d(v) < n$ forms the lower bound on the actual distance from $v$ to $r$ in the residual graph $G_f$ and if $d(v)\geq n$, then $d(v)-n$ is a lower bound on the actual distance from $r$ in the residual graph. 
\item A relabeling operation is initiated when a vertex with excess flow has a label less than or equal to that of the neighbouring vertex. Once relabeling is done, it can initiate a push operation. So one can think labels to denote the potential level, where flow can occur from a region of higher potential to a region of lower potential.

\item A vertex $v$ is defined as active $v \in V - \{s,r\}, d(v) < \infty$, and $e(v)>0$. The maximum-flow algorithm is initialized with preflow value $f$, which is summation of the edge capacities of all edges incident from the source vertex $s$ and rest all other edges have zero flow.  
\end{itemize}

For our distributed algorithm, \textit{HushRelay}, the basic operations is \textit{Push} and \textit{Relabel}, with all the nodes acting as individual processing unit in parallel. The network model considered for payment channel network is asynchronous. 
Synchronization across all the nodes is achieved via use of \textit{acknowledgements} \cite{tarjan}. A vertex $v$ tries to push excess flow to one of its neighbouring vertex $w$ if and only if, as per the information maintained by $v$, label $d(v)=d(w)+1$. It first sends a request message with the information $(v,\delta,d(v),e(v))$. Vertex $w$ can either accept the push by sending an acknowledgement or it may reject it by sending a negative acknowledgement $(NAK)$. If $d(v)=d(w)+1$, then $w$ sends to $v$ a message of the form $(accept,w,\delta,d(w))$ and $v$ initiates the push. Otherwise, if $d(w)\geq d(v)$ or $d(v) < d(w)+1$, then it sends a message $(reject,w,\delta,d(w))$ where $d(w)$ is the updated distance label of $w$. A reject message will cause $v$ to update the value of $d(w)$. When a distance label of the vertex increases, it sends the information of new label to all its neighbouring nodes. 

As seen in \textit{Push Relabel} algorithm \cite{tarjan}, the label initially set for source and sink node reveals the identity of payer and payee. To obfuscate their identity from other intermediate nodes in the network, we use a dummy source vertex $s'$ for node $s$ and a dummy sink vertex $r'$ for node $r$. Note that the existence of dummy node is known only by the source and sink. In the initialization phase of \textit{HushRelay}, a directed virtual edge from $s'$ to $s$ and from $r$ to $r'$ is established. Since $s',r'$ are virtual entities, introduction of edge $(s',s)$ and $(r,r')$ is not recorded in the blockchain. The capacity is initialized to $c(s',s)=val, c(r,r')=val$  and the label is set as $d(s')=n+2, d(s)=0, d(r)=0$ and $d(r')=0$. The flow $f(s',s)$ is set to $val$, $f(r,r')=0$, excess flow $e(s)=val$, $e(r)=e(r')=0$. For all vertices $v \in V-\{s,r\}$, $d(v)=0,e(v)=0$, $f(w,v)=0, (w,v) \in E, w,v \in V-\{s,r\}$. We mention the procedure of \textit{Push}, \textit{Push-request} and \textit{Relabel} for a vertex in Procedure \ref{algo:v4}, \ref{algo:vpr} and \ref{algo:v5} respectively.
\begin{proc}[!ht]
    \SetKwInOut{Input}{Input}
    \SetKwInOut{Output}{Output}

     \Input{ Active vertex $v \in V, e(v)>0$, vertices $w: (v,w) \in E$ }
      
    \caption{Push(v,w,d)}
        \label{algo:v4}
     \If{ $v\neq r'$ and $v \neq s'$}
     {
         Set $find\_neighbour$=0 \\
         \While{neighbour $w$ of $v$ : $e(v)>0, r_f(v,w)>0 \ and \ d(w) < d(v)$}
         {
         	$v$ generates a push of the value $\delta=min(e(v),r_f(v,w)$ \\
         	$f(v,w)=f(v,w)+\delta$\\
         	$e(v)=e(v)-\delta$\\
         	$r_f(v,w)=c(v,w)-f(v,w)$\\
         	Send \textit{Push-request(w,v,$f(v,w),\delta$)} to node $w$ \\
         	\If{\textit{NAK} received}
         	{
         	Update information $d(w)$ \\
         	$f(v,w)=f(v,w)-\delta$\\
         	$e(v)=e(v)+\delta$\\
         	$r_f(v,w)=c(v,w)+f(v,w)$\\
         	}
         	\Else
         	{
         	 Set $find\_neighbour=1$\\
         	 
         	}
         }
         \If{$find\_neighbour=0$}
          {
             Call \textit{Relabel} function.
          }
     }
\end{proc}


\begin{proc}[!ht]
    \SetKwInOut{Input}{Input}
    \SetKwInOut{Output}{Output}

     \Input{ Active vertex $w \in V, e(w)>0$, vertex $v: (w,v) \in E$ }
      
    \caption{Push-request(v,w,$f(w,v)$,$\delta$)}
        \label{algo:vpr}
     \If{ $d(v)<d(w)$}
     {
         $r_f(v,w)=f(w,v)$\\
         $e(v)=e(v)+\delta$ \\
         \If{$v\neq t'$ and $v \neq s'$}
         {
         	  send \textit{Push} request to its neighbouring nodes.\\

         }
         Send \textit{push request accepted} message to node $w$.
     }
     \Else{
          Send \textit{negative acknowledgement (NAK)} message and current value of $d(v)$ to node $w$.
     }
\end{proc}


\begin{proc}[!ht]
    \SetKwInOut{Input}{Input}
    \SetKwInOut{Output}{Output}

     \Input{ Active vertex $v \in V, e(v)>0$, vertices $w: (v,w) \in E, r_f(v,w)>0, d(v)\leq d(w)$ }
      
    \caption{Relabel(v,w,d)}
        \label{algo:v5}
     \begin{enumerate}
     \item Update $d(v)=min(d(w) ,(v,w) \in E) +1$
     \item Inform all the neighbours of vertex $v$ about the updated label $d(v)$.
     \end{enumerate}
\end{proc}

The algorithm terminates when there are no active vertex left (except the dummy source and dummy sink) in the graph. The number of messages exchanged (\textit{for push request, push accepted/NAK, height updation}) is also bounded. The communication complexity, termination condition of the algorithm and an upper bound on the $d$ value of a node in the given graph is stated in \cite{thuy2005distributed}. $\mathcal{O}(n^2m)$ messages are exchanged in the asynchronous implementation with $\mathcal{O}(n^2)$ runtime. The overhead lies in the interprocessor communication between a vertex and its neighbours.

\begin{example}
Consider a network given in Fig. \ref{fig:example1}. Sender $S$ intends to make a payment of 15 units to receiver R. Dummy vertices $S'$ and $R'$ is added to the network with edges $(S',S)$ and $(R,R')$. The edge capacities are as follows : $c(S',S)=15, c(S,A)=10, c(S,B)=10, c(A,C)=10, c(B,C)=15, c(C,R)=20$ and $c(R,R')=15$. \textit{HushRelay} is implemented on the network to obtain a feasible flow of value 15 from source node S to sink node R. The initial state is given in Fig. \ref{fig:example1} (a). 
\begin{figure}[!ht]%
    \centering
    \label{img1}

    \subfloat[Initial state]{{\includegraphics[width=6.3cm]{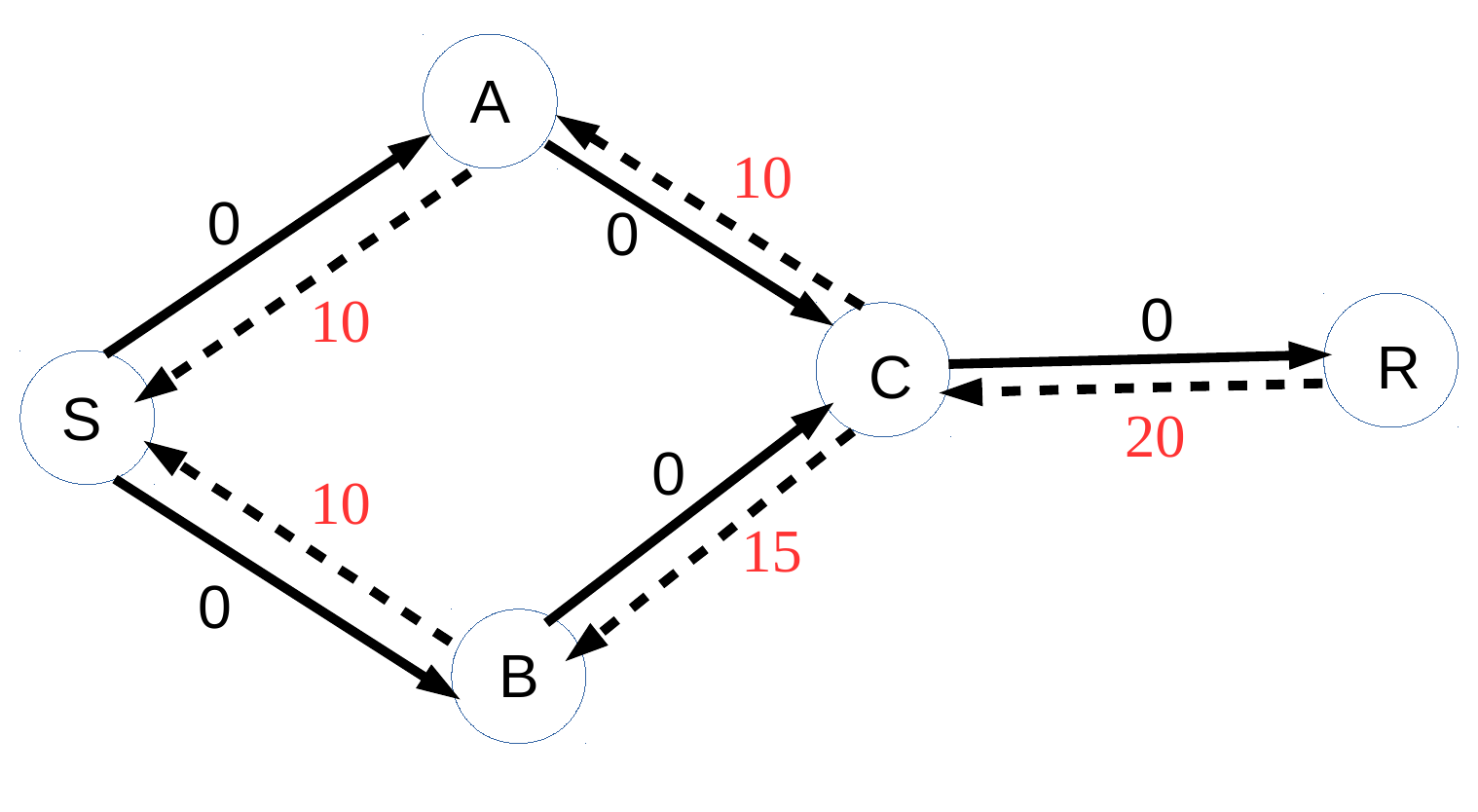} }}%
    \end{figure}

In the initialization phase, each nodes is assigned a label of 0 except dummy vertex $S'$ where d(S') is the count of the number of nodes (except S') in the network. As given in Fig. \ref{fig:example1} (b), S' sends a push request of 15 units to S.  

\begin{figure}[!ht]\ContinuedFloat

    \centering 
               \subfloat[state]{{\includegraphics[width=6.3cm]{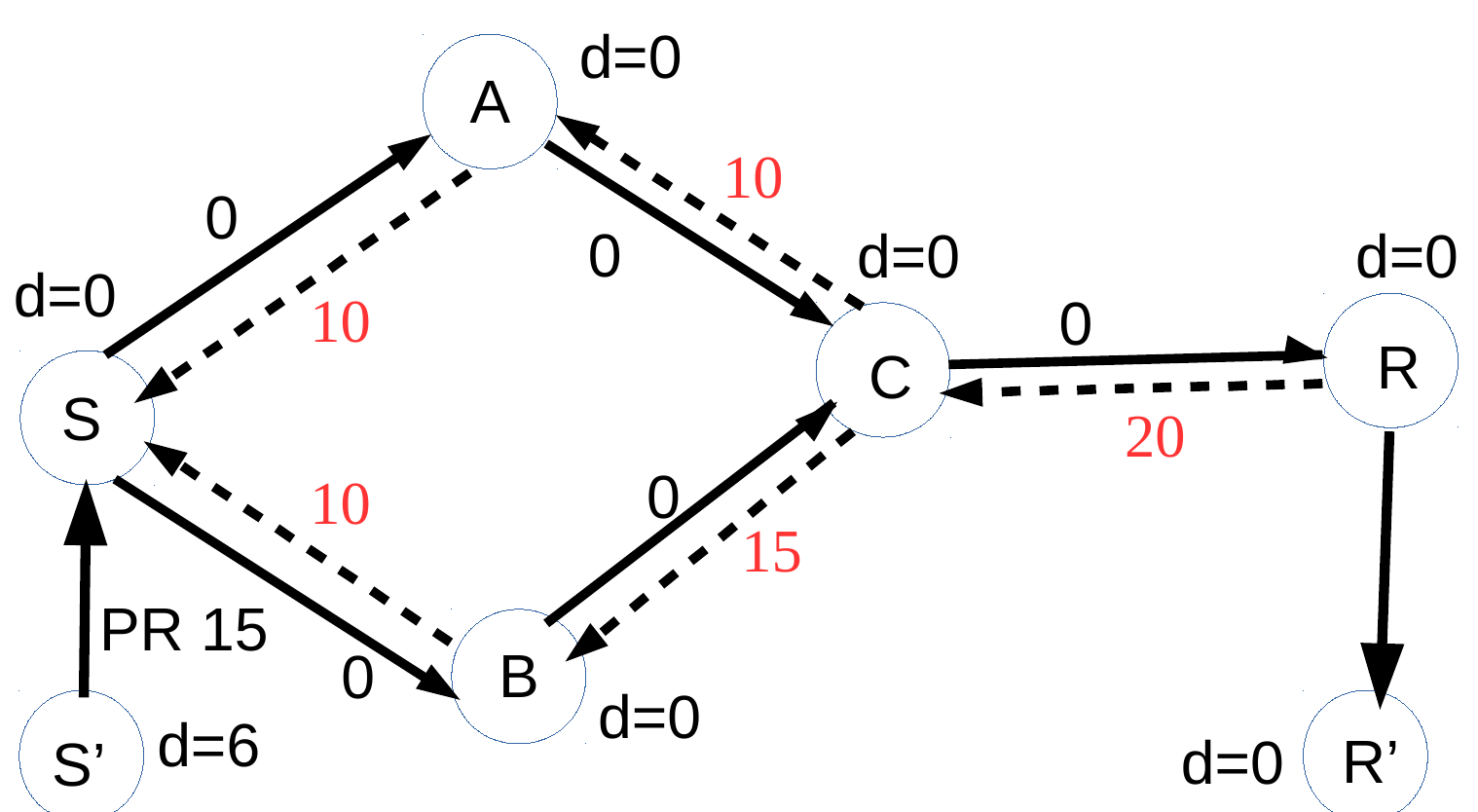} }}%
    \end{figure}
 In Fig. \ref{fig:example1} (c), S accepts the push request as $d(S)<d(S')$ and an excess flow of 15 units is assigned to S. S changes its label calling \emph{relabel} function and changes $d=1$. Now S sends a push request of 10 units to A, bounded by the capacity of payment channel SA.     
\begin{figure}[!ht]\ContinuedFloat
    \centering

        \subfloat[state]{{\includegraphics[width=6.3cm]{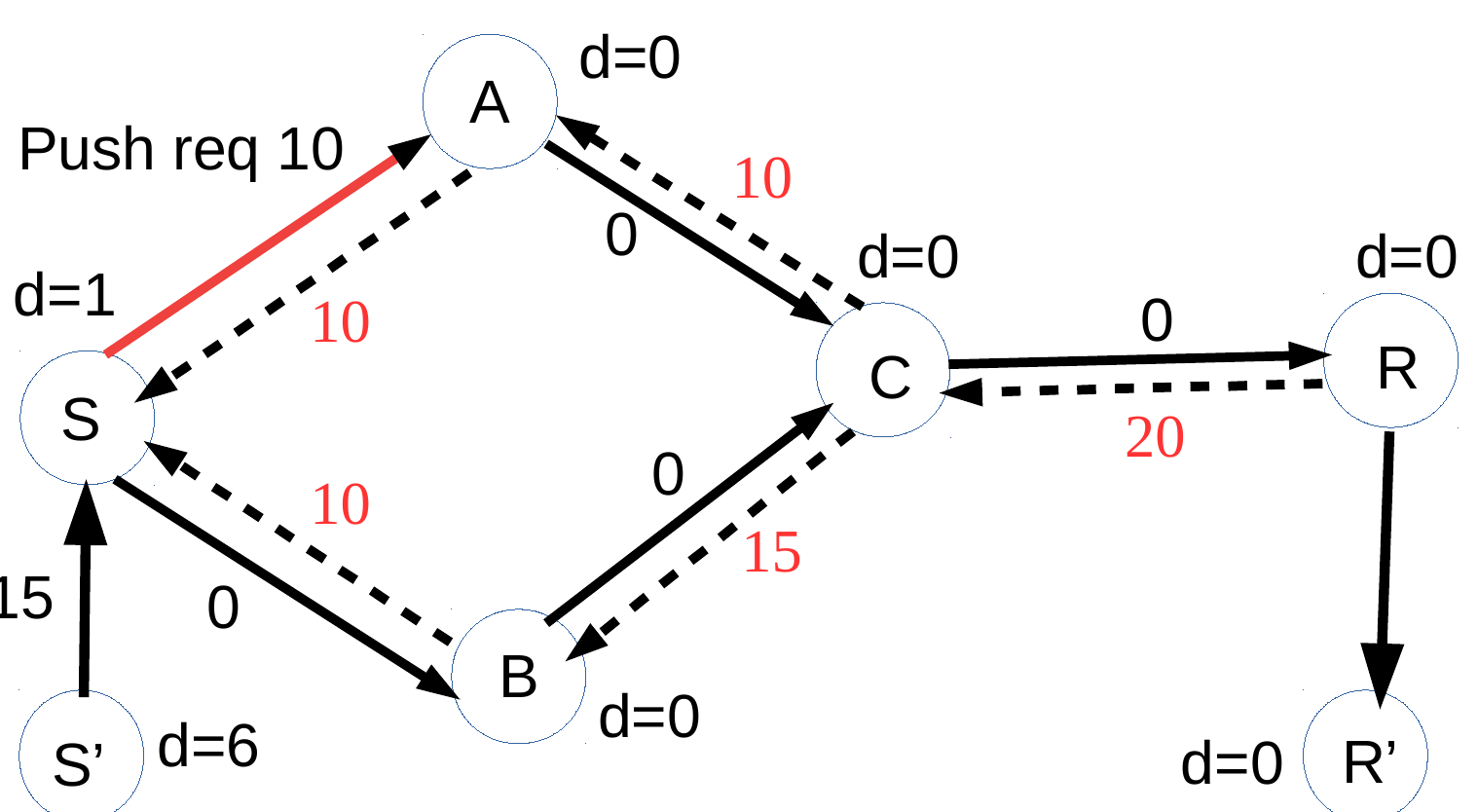} }}%
    \end{figure}
 \\In Fig. \ref{fig:example1} (d), A accepts the push request as $d(A)<d(S)$ and gets an excess flow of 5 units. $d(A)$ is changed to 1. S still has an excess flow of 5. It sends a push request to B. Simultaneously A sends a push request of 10 units to C. 
\begin{figure}[!ht]\ContinuedFloat
    \centering

    \subfloat[state]{{\includegraphics[width=6.3cm]{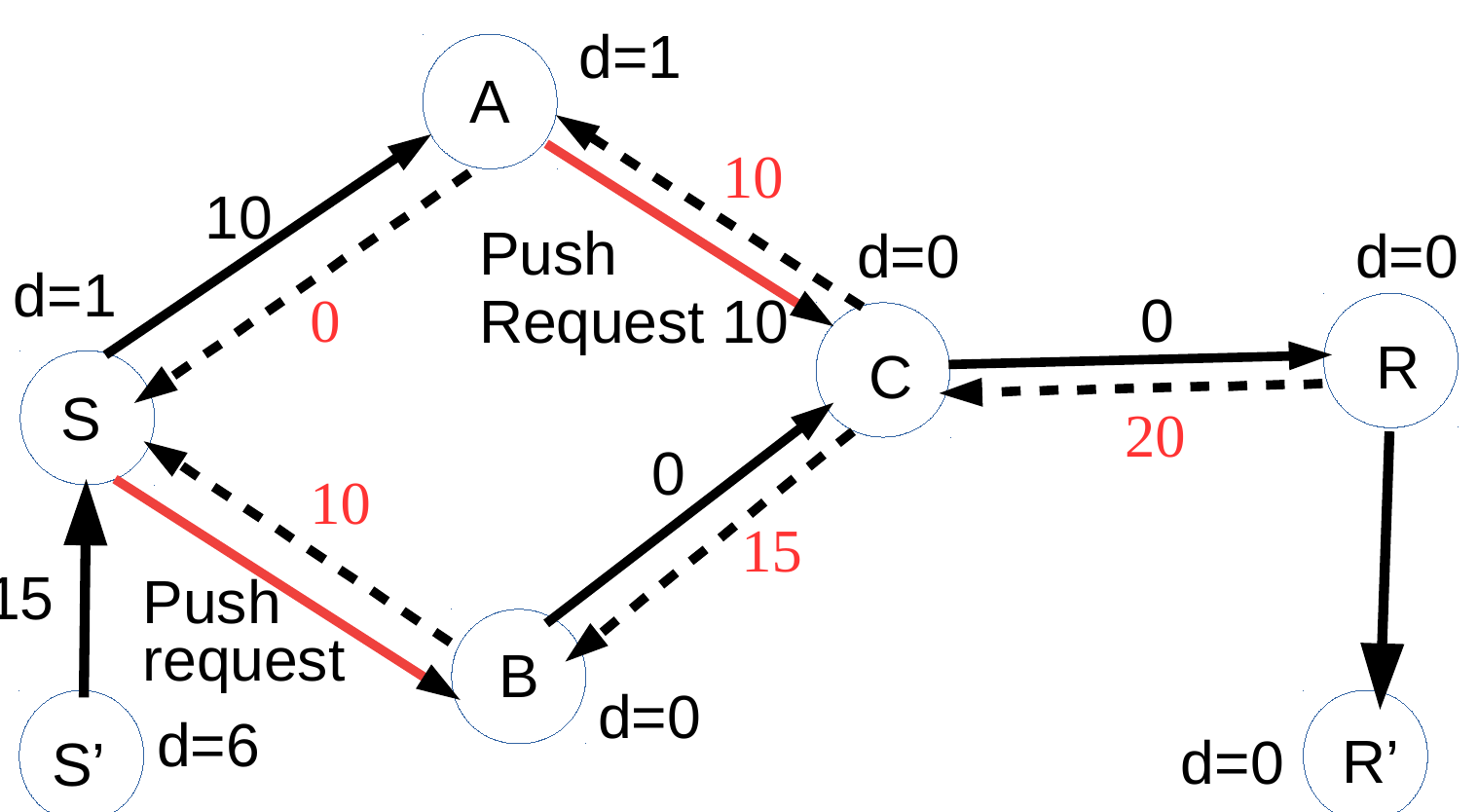} }}%
            
\end{figure}
In Fig. \ref{fig:example1} (e), B accepts the push request as $d(B)<d(S)$ and has excess flow of 5 units. $d(B)$ is changed to 1. Similarly C accepts push request as $d(C)<d(A)$ and has an excess flow of 10 units. $d(C)$ is changed to 1. B sends a push request to C and C sends a push request to R. In Fig. \ref{fig:example1} (f), upon receipt of request from B, C finds that $d(C)=d(B)$ and hence it sends a NAK to B. R accepts the push request from C and gets an excess flow of 10. It changes its label to $d(R)=1$ and consequently, it sends a push request of 10 units to R'. In Fig. \ref{fig:example1} (g), B changes its label to $d(B)=2$. R' accepts the push request.
\begin{figure}[!ht]\ContinuedFloat
    \centering
    \subfloat[state]{{\includegraphics[width=6.3cm]{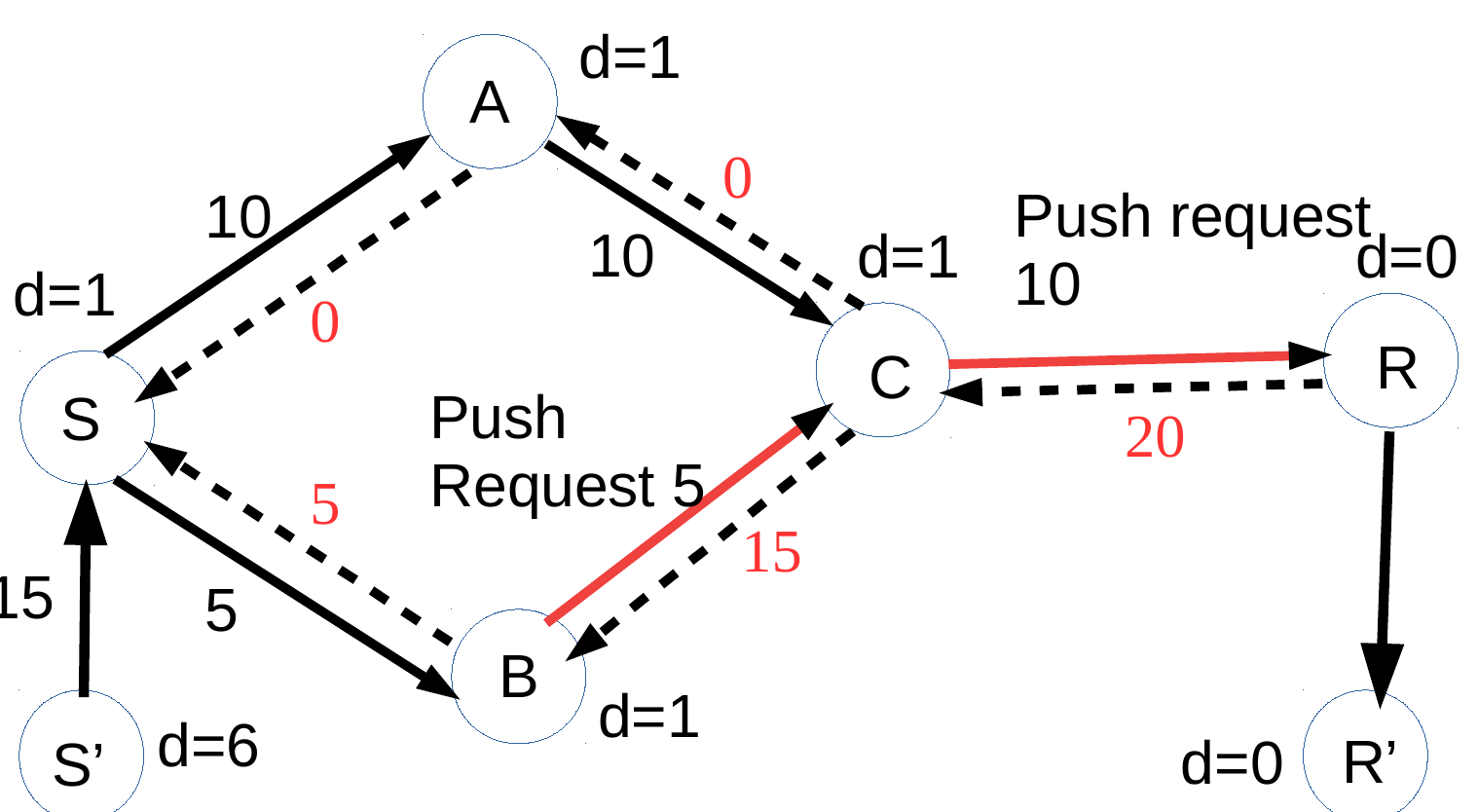} }}%
 \end{figure}

 \begin{figure}[!ht]\ContinuedFloat
    \centering
            \subfloat[state]{{\includegraphics[width=6.3cm]{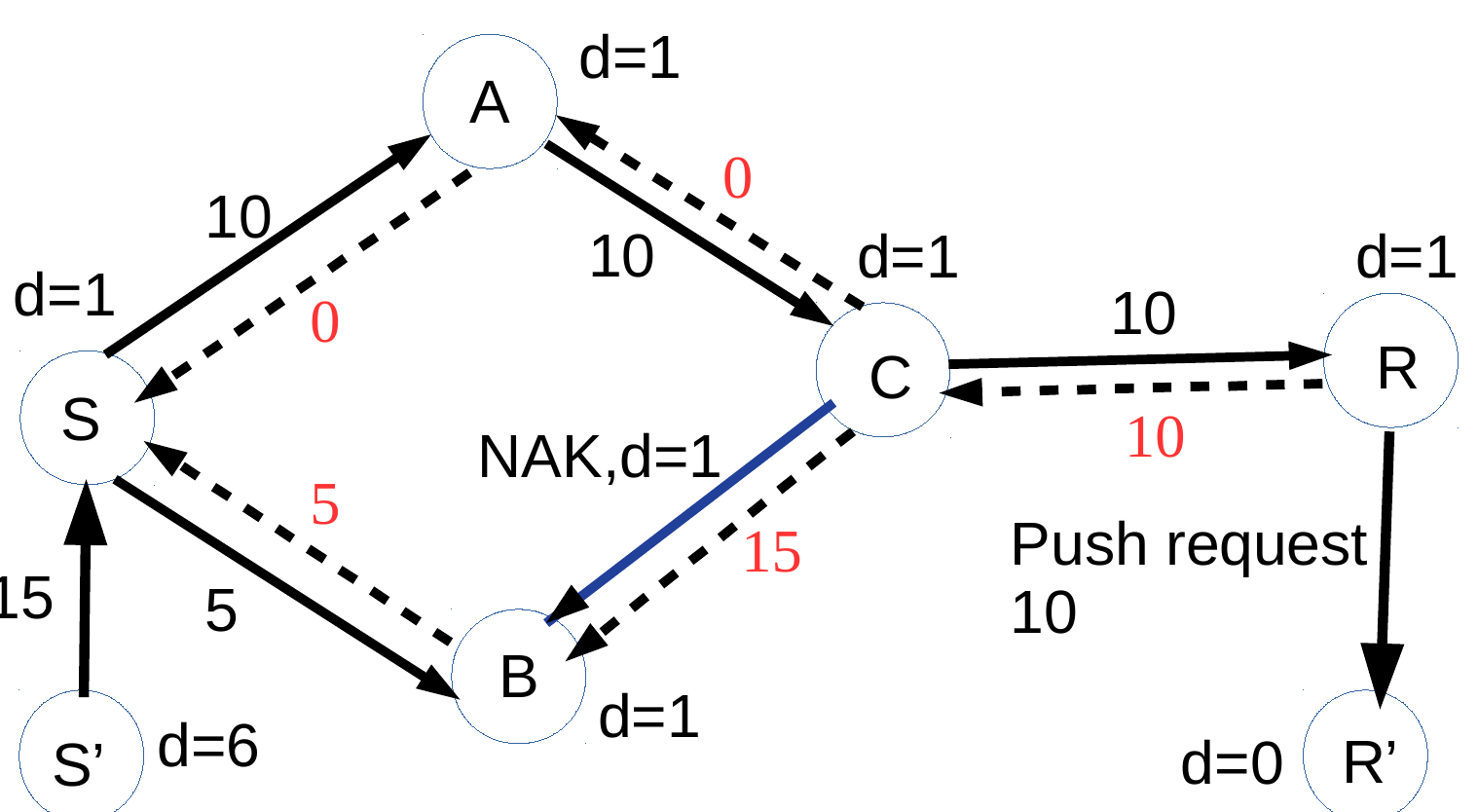} }}%
\end{figure}

 \begin{figure}[!ht]\ContinuedFloat
    \centering

    \subfloat[Initial state]{{\includegraphics[width=6.3cm]{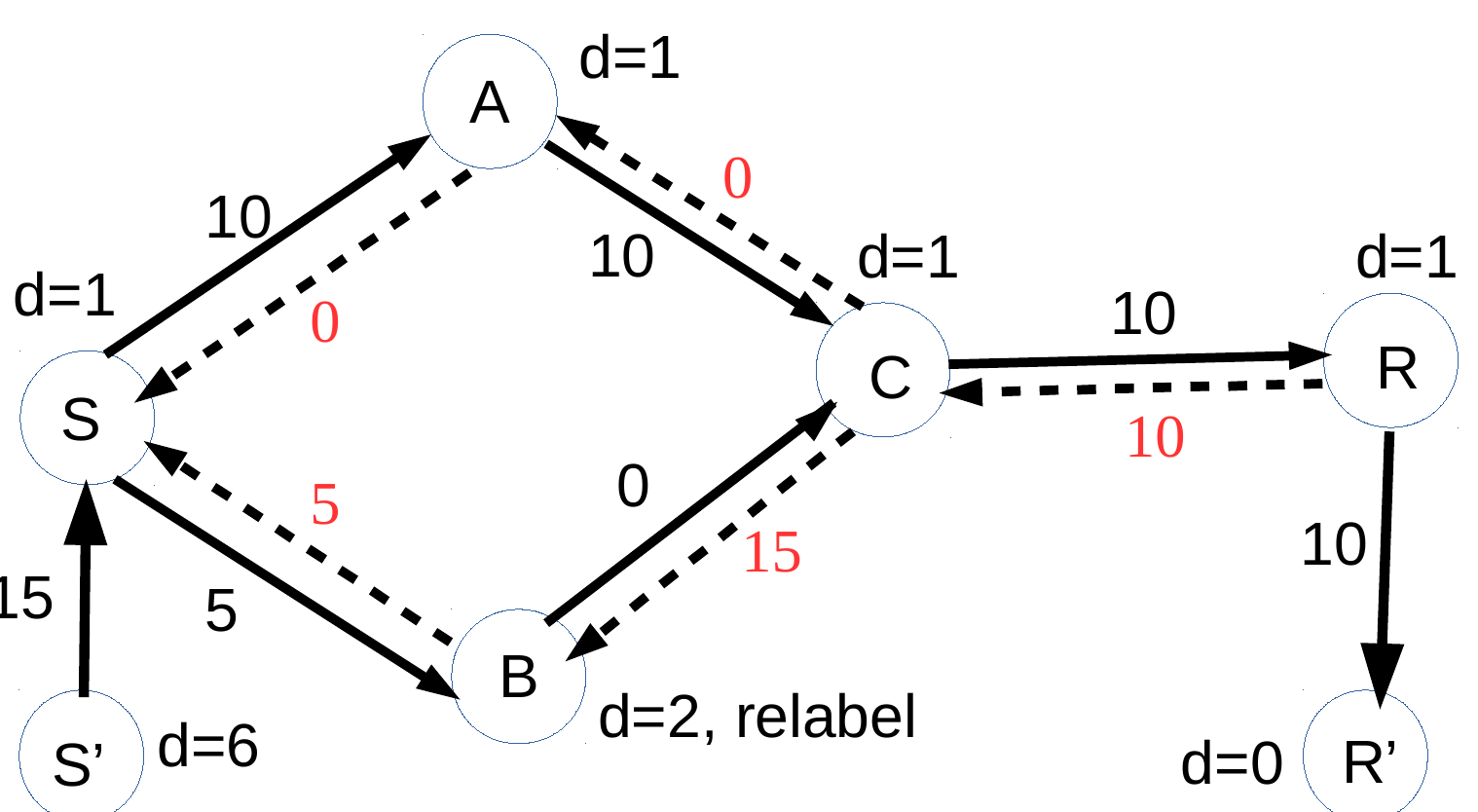} }}%
\end{figure}
In Fig. \ref{fig:example1} (h), B sends a push request of 5 units to C.

 \begin{figure}[!ht]\ContinuedFloat
    \centering
            \subfloat[Initial state]{{\includegraphics[width=6.3cm]{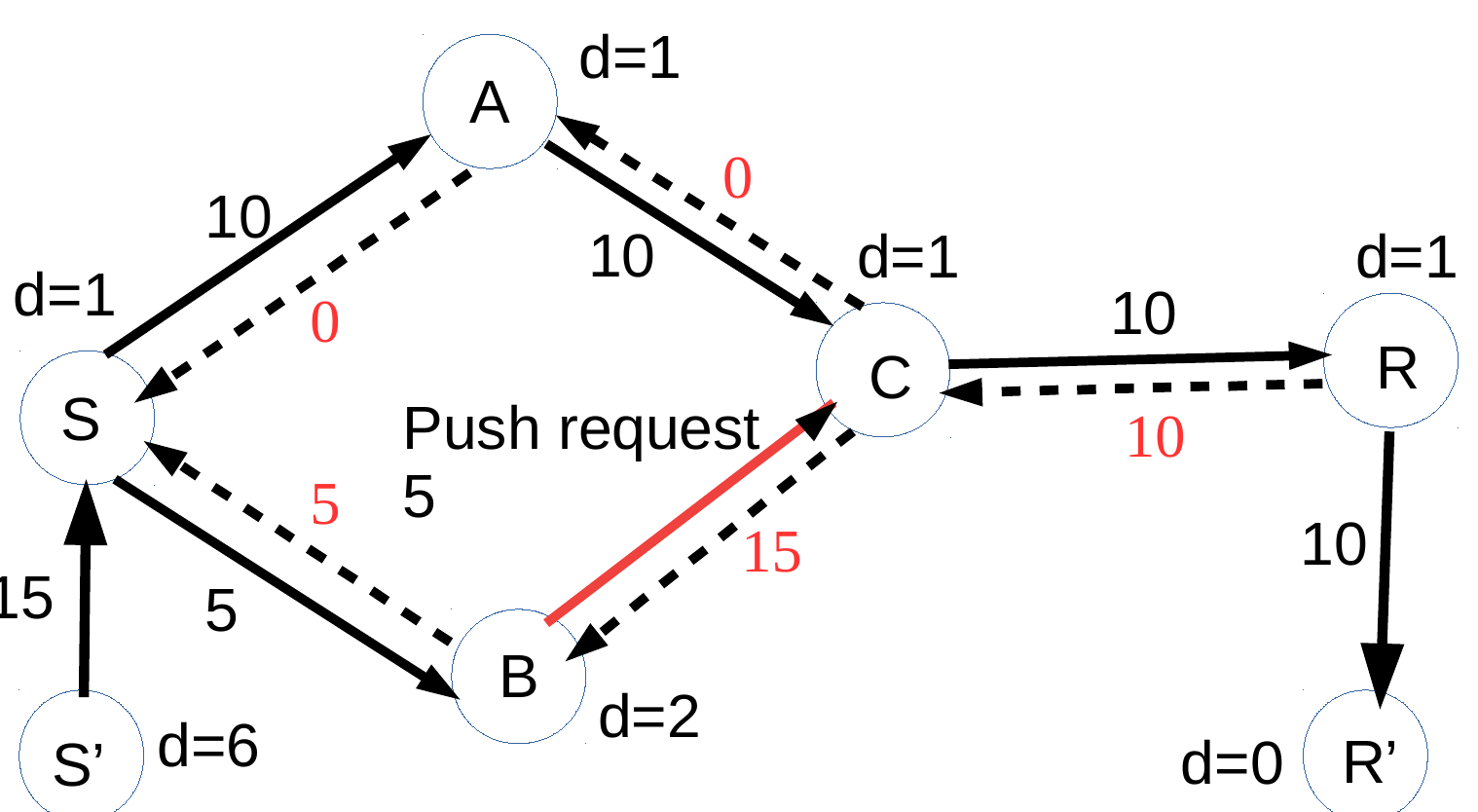} }}%
\end{figure}
 In Fig. \ref{fig:example1} (i), C accepts the push request and changes it label to 1. It sends a push request of 5 units to R.

 \begin{figure}[!ht]\ContinuedFloat
    \centering
        \subfloat[Initial state]{{\includegraphics[width=6.3cm]{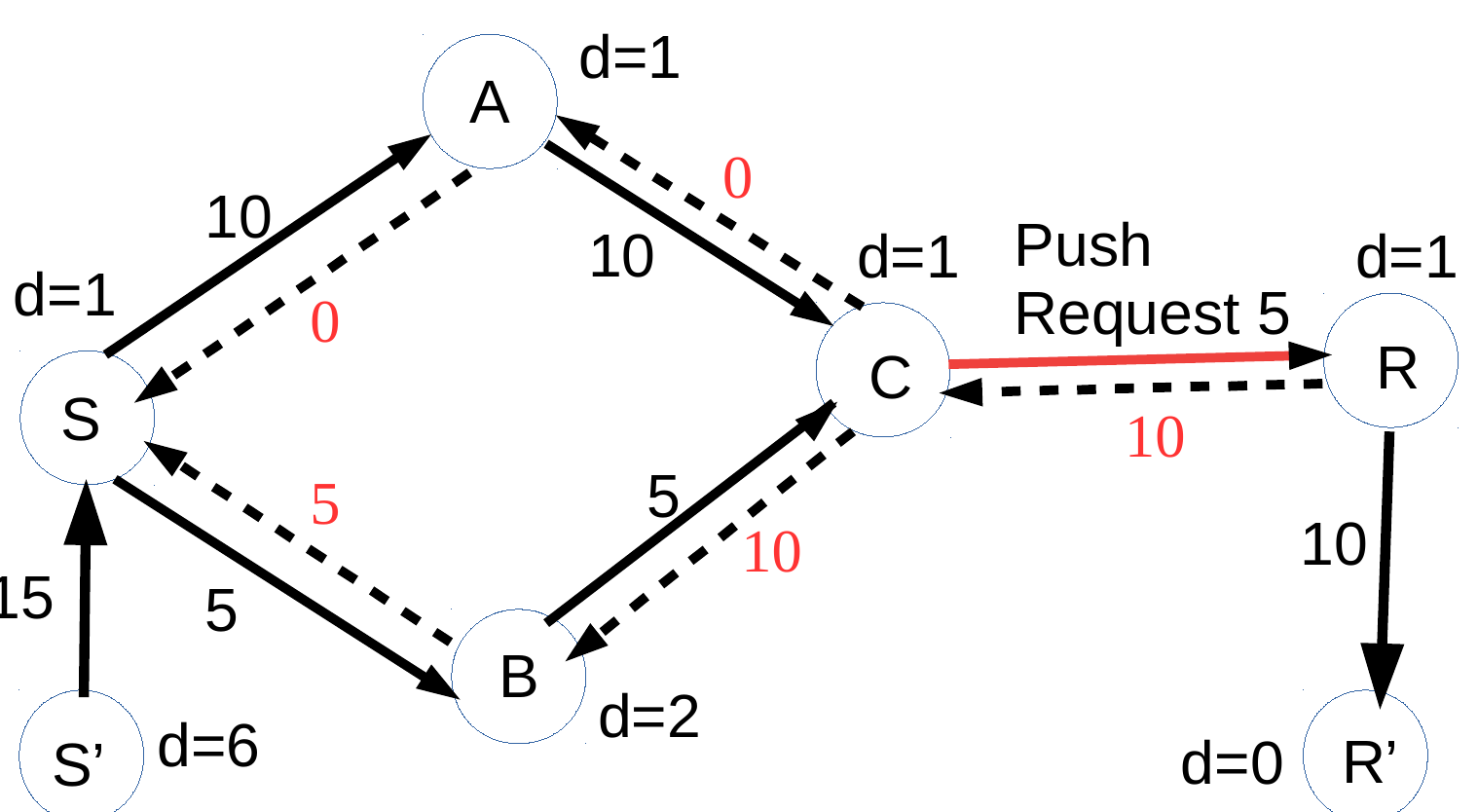} }}%
    \end{figure}

 In Fig. \ref{fig:example1} (j), R finds that $d(R)=d(C)$ and hence it sends a NAK to C. 

 \begin{figure}[!ht]\ContinuedFloat
    \centering
            \subfloat[Initial state]{{\includegraphics[width=6.1cm]{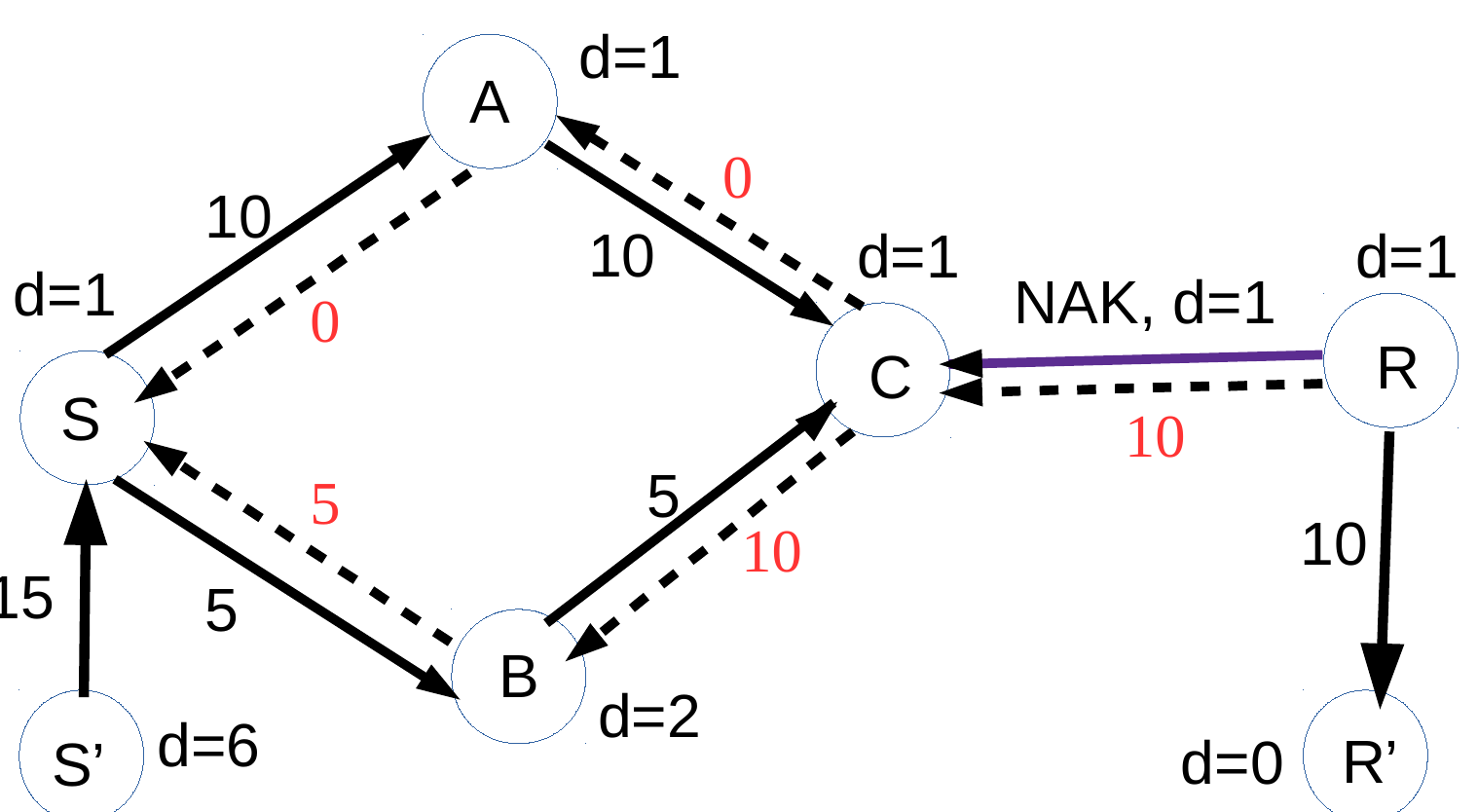} }}%
        \end{figure}
In Fig. \ref{fig:example1} (k), C undergoes a relabeling operation and $d(C)$ is changed to 2.
\begin{figure}[!ht]\ContinuedFloat
    \centering
            \subfloat[Initial state]{{\includegraphics[width=6.3cm]{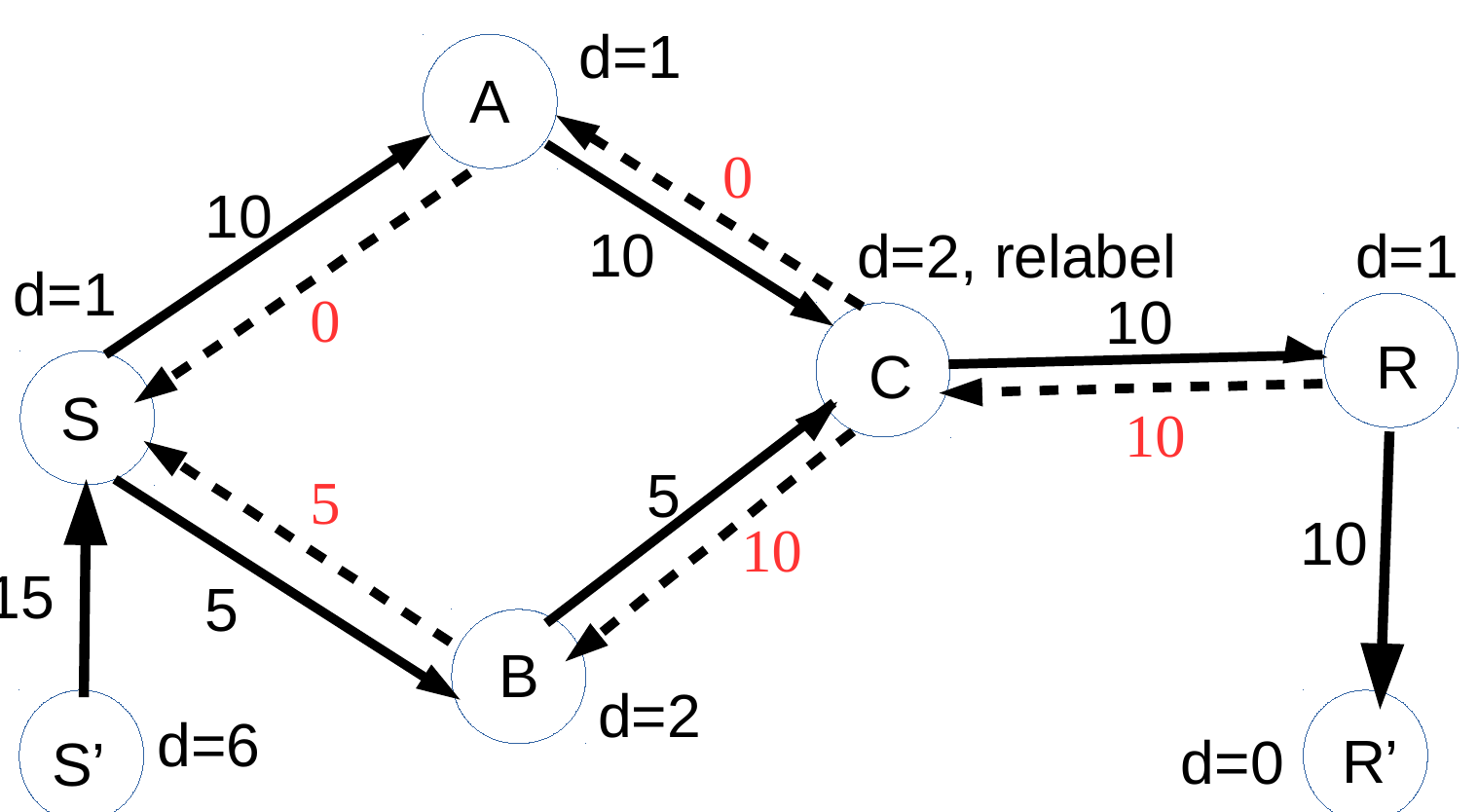} }}%
\end{figure}
   In Fig. \ref{fig:example1} (l), C again sends a push request of 5 units to R. 
\begin{figure}[!ht]\ContinuedFloat
    \centering

        \subfloat[Initial state]{{\includegraphics[width=6.3cm]{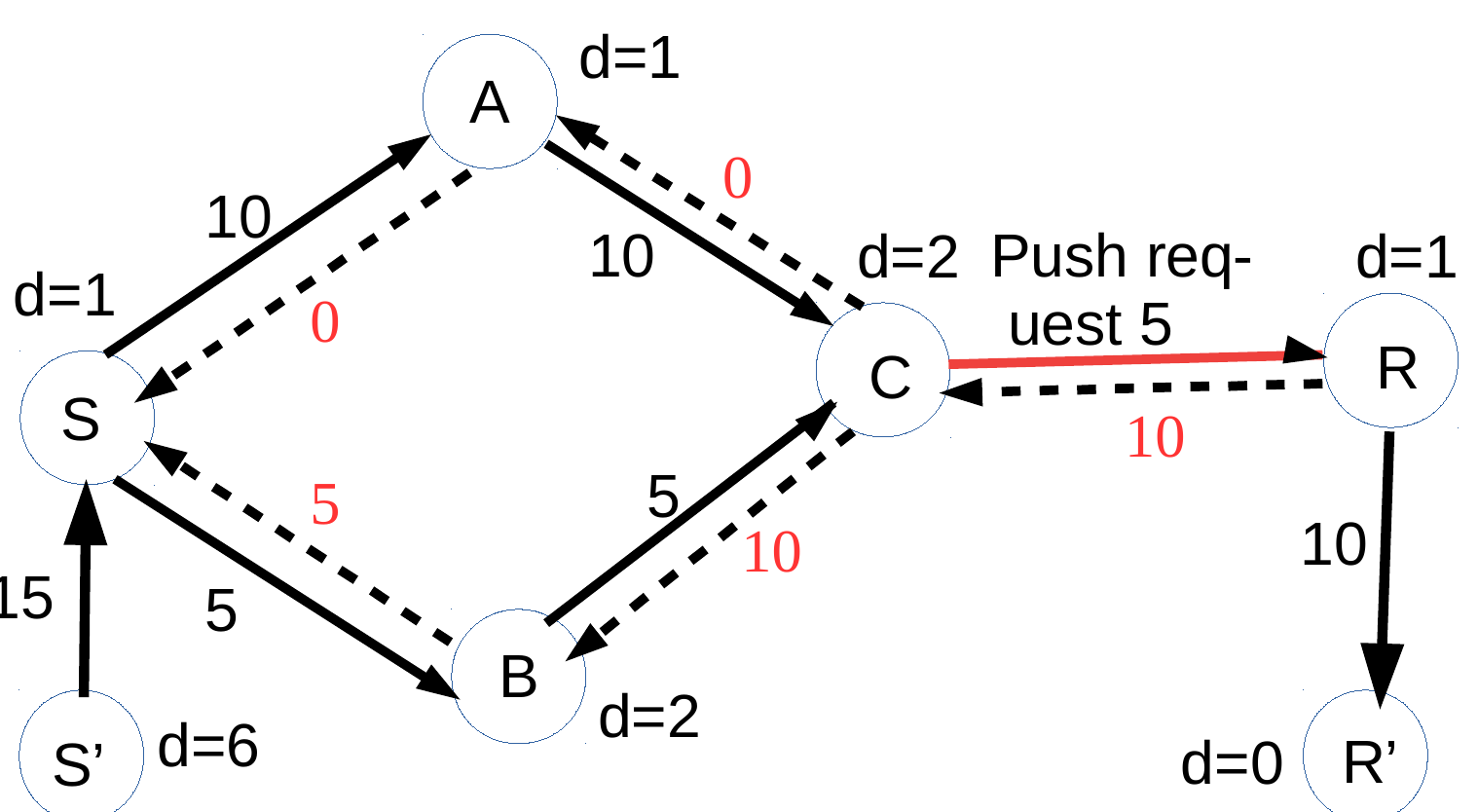} }}%
\end{figure}

In Fig. \ref{fig:example1} (m), R accepts the push request and it sends a push request of 5 units to R'.
\begin{figure}[!ht]\ContinuedFloat
    \centering

  \subfloat[Initial state]{{\includegraphics[width=6.1cm]{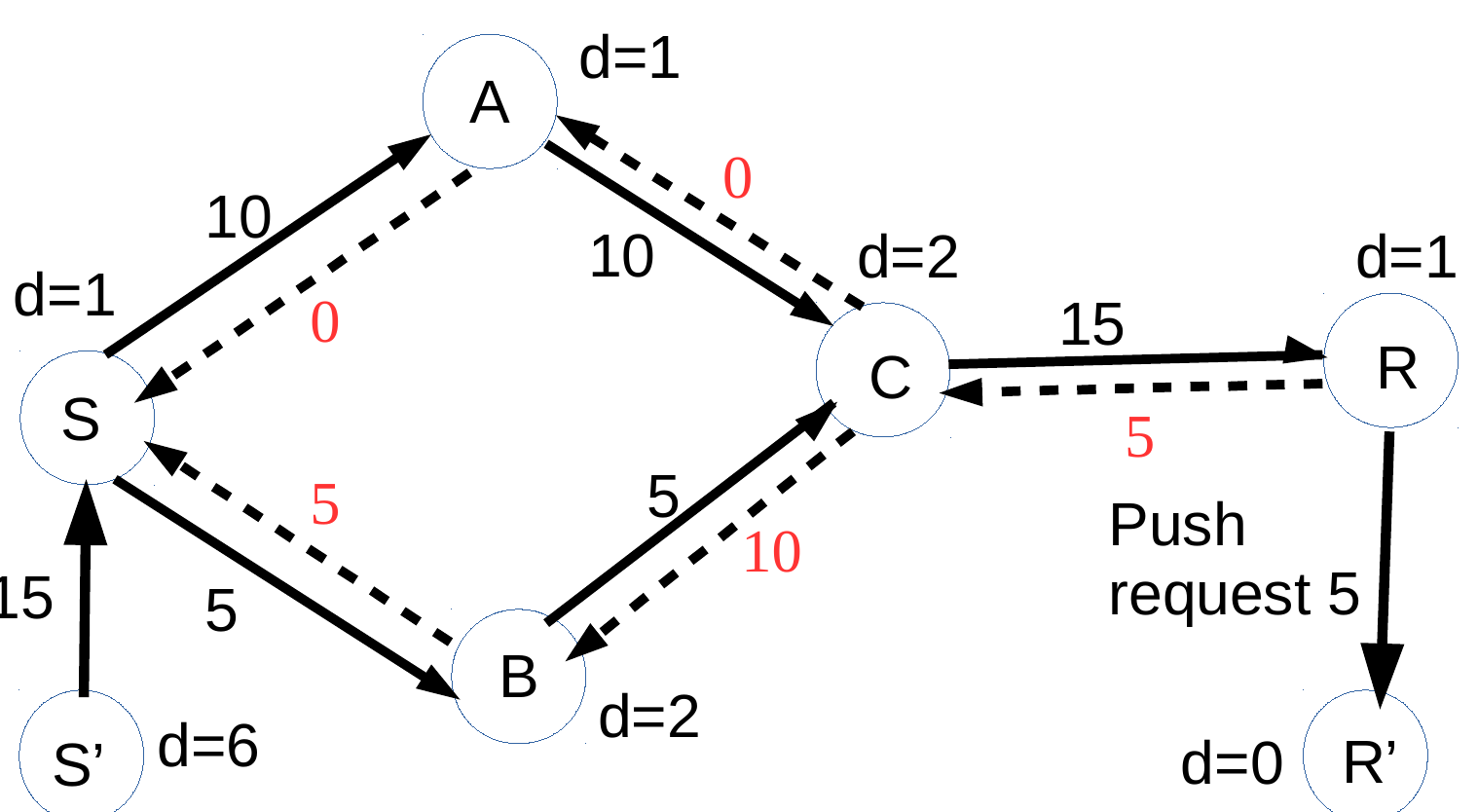} }}%
\end{figure}
   In Fig. \ref{fig:example1} (n), the algorithm terminates transferring 15 units to R'
\begin{figure}[!ht]\ContinuedFloat
    \centering

\subfloat[After addition of dummy vertices S' and R', execution of the routing algorithm]{{\includegraphics[width=6.1cm]{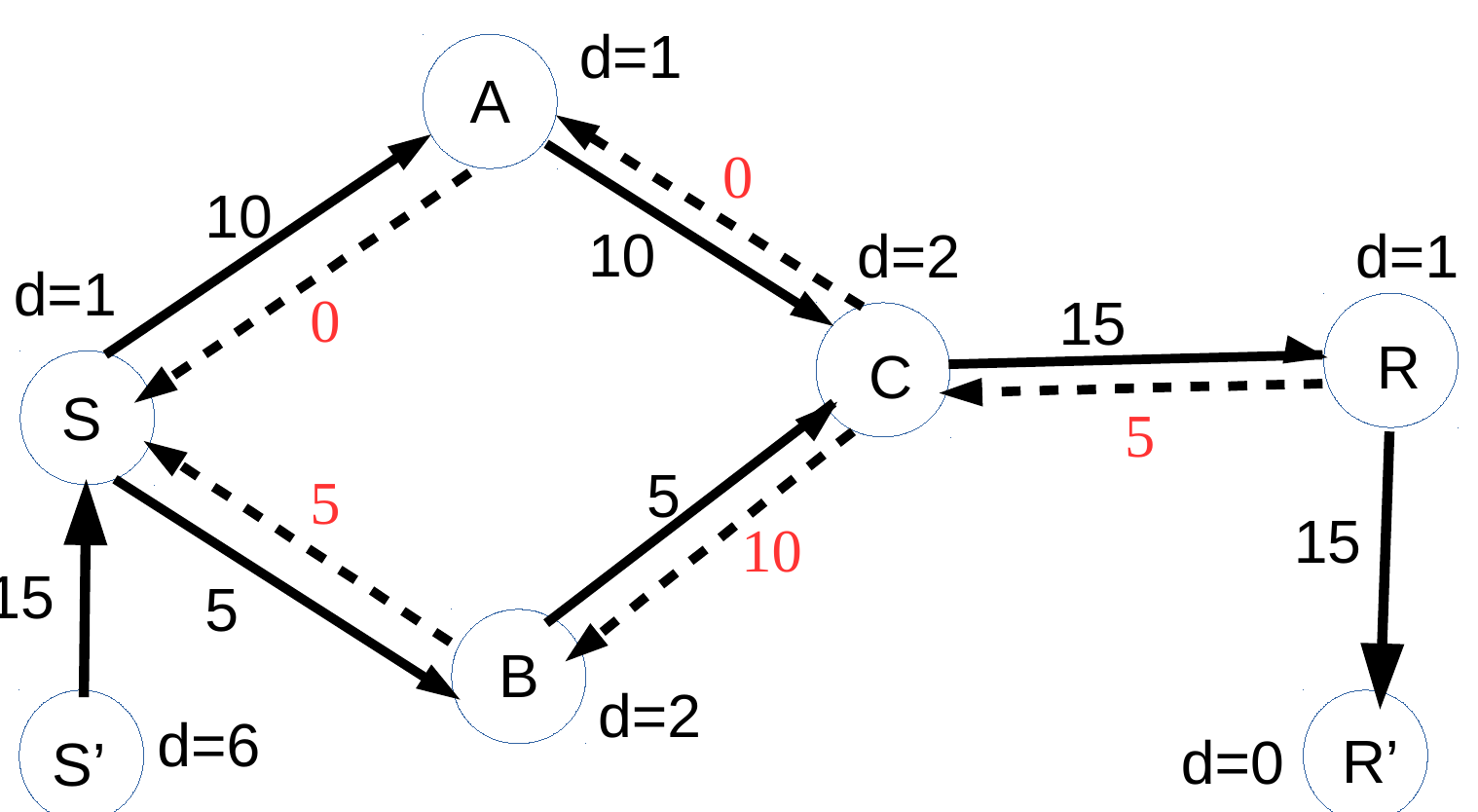} }}%

\caption{Execution of HushRelay}%
\label{fig:example1}%
\end{figure}
\end{example}

\subsection{Proof of correctness of the HushRelay}
\label{correct}
We state the following lemmas which justifies the correctness of our routing algorithm.
\begin{lemma}
\label{lm1}
If $val \leq \textrm{maximum flow}$ in $G$, then $s$ can successfully transfer funds to $r$.
\end{lemma}
\textbf{Proof of Lemma \ref{lm1}}. Given that $\tilde{d} < \textrm{maximum flow}$, let us assume that transaction from $s$ to $r$ fails due to non existence of sufficient capacity from sender to receiver. Now we execute the distributed push relabel algorithm which will return the maximum flow in the graph. Let us denote it by $f_m$. But our algorithm was able to find augmenting path for flow till $\tilde{d} - \gamma$, where $\gamma >0$ is an integral value, since our transaction failed. If there exists no more augmenting paths, then $f_m = \tilde{d} - \gamma$, which implies $f_m < \tilde{d}$. This contradicts the fact that $f_m > \tilde{d}$. Hence our assumption was wrong.     

\begin{lemma}
\label{lm2}
For $ v \in V\setminus \{s',t'\}, e(v)=0$ on termination.
\end{lemma}

\textbf{Proof of Lemma \ref{lm2}}. Assume that there exist one vertex $\hat{v} \in V\setminus \{s',t'\} : e(\hat{v})>0$ after termination. But since termination condition has been reached, it means there vertex $s$ is not reachable from this vertex $\hat{v}$. Let set of vertices not reachable from $\hat{v}$ be denoted by $V'$ and those reachable from $\hat{v}$ be $V-V'$.
\begin{equation}
\begin{matrix}

e(v)=\Sigma_{k \in V, (v,k) \in E} f(k,v) \\
    = \Sigma_{k \in V',(v,k) \in E} f(k,v) + \Sigma_{k \in V-V',(v,k) \in E} f(k,v) \\
    = \Sigma_{k \in V',(v,k) \in E} f(k,v) \\ (\because \Sigma_{k \in V-V',(v,k) \in E} f(k,v)=0 , \textrm{flow conservation constraint})
\end{matrix}
\end{equation}   
But since $e(v)>0$, then $\Sigma_{k \in V',(v,k) \in E} f(k,v)>0$, which means there still exists some augmenting path from $s$ to $v$ ($s \in V'$). Hence it contradicts the assumption of termination. 

\begin{lemma}
\label{lm3}
For all edges $(v,w) \in E, v,w \in V$, $f(v,w)\leq c(v,w)$.
\end{lemma}

\textbf{Proof of Lemma \ref{lm3}}. In the algorithm \textit{Push}, the flow value $\delta$ for a given edge $(v,w) \in E$, the flow value $f(v,w)$ from vertex $v$ to vertex $w$ is decided by $\min(e(v),r_f(v,w))$. Since $r_f(v,w)=c(v,w)-f(v,w)$,  $f(v,w)\leq c(v,w)$ and $e(v) \leq \tilde{d}$. Flow value will be bounded by $\tilde{d}$, if $\tilde{d} < c(v,w)$ or $c(v,w)$ otherwise.

\subsubsection{\textbf{Propagating the flow information to source node}}
\label{hide}
Each edge $e \in E$ involved in transfer of payment from source to sink will generate temporary key $k_e$ for encrypting the flow message to be propagated back to the source node. The sink node, on termination, generates a key $k_{sink}$ as well some random message $rm$, equivalent to size of the packet or its multiple. Each such packet contains flow information which is shared with a predecessor node. It constructs a message $m'$ containing the information of identity of preceding vertex $w$, the non negative flow  $f_{wv}>0$ along with key $k_{wv}$. It encrypts the packet with $k_{sink}$, $E'=Enc_{k_{sink}}(w,f_{wv},k_{wv})$, and concatenates the randomly generated message $rm$ with the encrypted packet to construct message $E'||rm$. It shares this information with $w$. If $w$ is honest, it will construct a similar message $m'$, for its neighbour say $u$ containing the identity of $u$, flow $f_{uw}$, key $k_{uw}$. It is encrypted with $k_{wv}$ to get $E''$. The encrypted message is concatenated with the message received from its successor i.e. $E''||E'||rm$. This continues till all the packets reach the source vertex $s$. The sink vertex shares $k_{sink}$ and set of randomly generated message $rm$ with source vertex $s$ via secure communication channel. $s$ discards $rm$ from the received message and starts decrypting, beginning with the message encrypted by sink. On decryption, it retrieves the flow information, identity of vertex and key with which it will decrypt the next encrypted packet. All duplicate information on flow is discarded and the remaining one is used for reconstructing the flow across the network. This is the routing information denoted by $\mathcal{P}$.


 \begin{table*}[!ht]
\begin{center}

 \caption{SpeedyMurmur vs HushRelay - Performance Analysis on Real Instances }
   \label{tab:1}
  \scalebox{0.9}[0.9]{
%
%


  \begin{tabular}{|p{3cm}|p{1cm}|p{1cm}|p{1cm}|p{1cm}|p{1cm}|p{1cm}|p{1cm}|p{1cm}|p{1cm}|p{1cm}|} 
  \hline
\multirow{4}{*}{Network/Algorithm} &
      \multicolumn{8}{c|}{SpeedyMurmur} &
      \multicolumn{2}{c|}{HushRelay} \\
      \cline{2-11}
 
     & \multicolumn{4}{c|}{Success Ratio} &\multicolumn{4}{c|}{Time taken}
     & Success Ratio &Time taken\\

 &
      \multicolumn{4}{c|}{Number of Landmarks} &
      \multicolumn{4}{c|}{Number of Landmarks} & &  \\
      \cline{2-9}
      &1 &2 &4 &6 &1 &2&4 &6 & &\\
\hline
Ripple Network  &0.38 &0.7 &0.92 &0.98 &1.66s &2.2s &3.23s &4.74s &1 &2.4s\\
\hline
Lightning Network  &0.42 &0.65 &0.83 &0.91 &0.61s &0.69s &0.83s &1.94s &0.99 &0.15s\\
\hline
 \end{tabular}

}
\end{center}
\end{table*}

\section{Performance Analysis of \textit{HushRelay}}
\label{6}
\subsubsection*{Experimental Setup}
In this section, we define the experimental setup. The code for \textit{HushRelay} is available in \cite{Code}. System configuration used is : \texttt{Intel Core i5-8250U CPU, Kabylake GT2 octa core processor, frequency 1.60 GHz}, OS : \textit{Ubuntu-18.04.1 LTS} (64 bit). The programming language used is C, compiler - gcc version 5.4.0 20160609. The library \textit{igraph} was used for generating random graphs of size ranging from 50 to 25000, based on Bar\'{a}basi-Albert model \cite{albert2002statistical}, \cite{barabasi2003scale}. Payment Channel Network follows the scale free network where certain nodes function as hub (like central banks), having higher degree compared to other nodes \cite{javarone2018bitcoin}. For implementing the cryptographic primitives, we use the library \textit{Libgcrypt}, version-1.8.4 \cite{libgcrypt}, which is based on code from GnuPG. 
\begin{figure}[h]
\centering
\subfloat[Success Ratio vs Number of Nodes]{{\includegraphics[width=8cm]{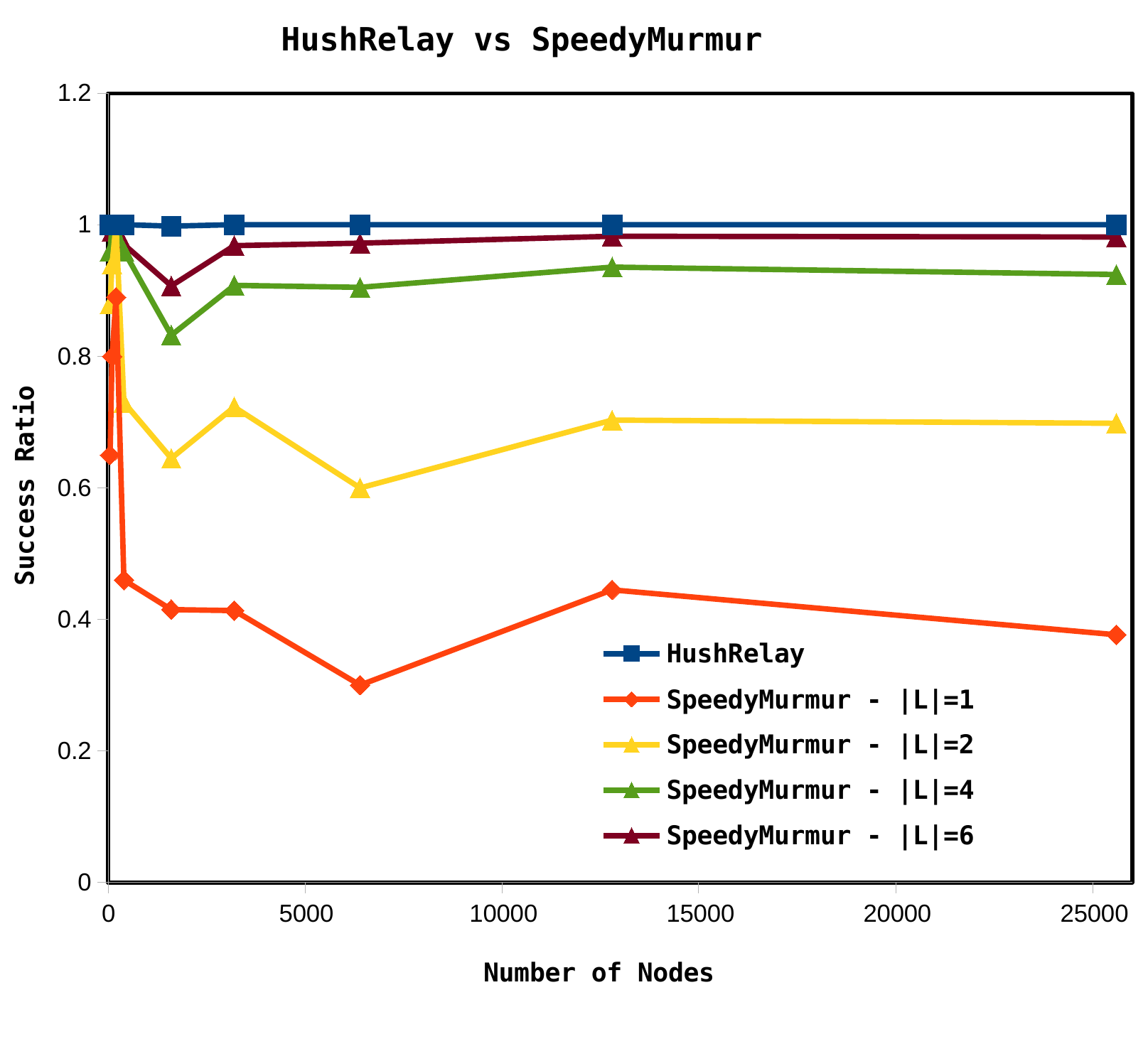}}}%
\qquad
\subfloat[Time To Route vs Number of Nodes]{{\includegraphics[width=8cm]{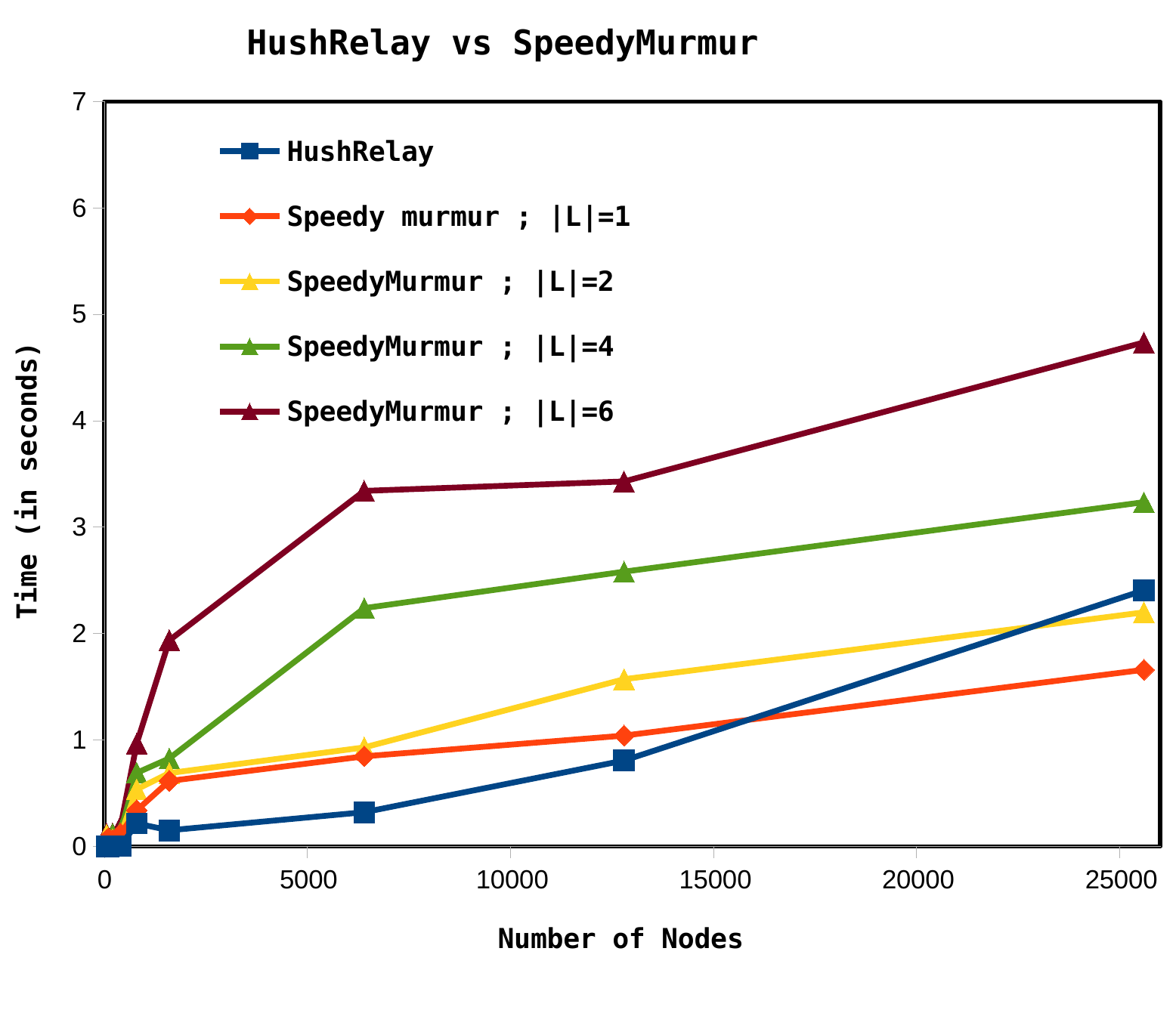} }}%
\qquad
\caption{Analysis of \textit{HushRelay} and \textit{SpeedyMurmur}}%
\label{fig:example2}%
\end{figure}
\subsubsection*{Evaluation}
Following metrics are used to compare the performance of the routing algorithm, \textit{HushRelay} with \textit{SpeedyMurmur} \cite{speedymurmur} 
\begin{itemize}[leftmargin=*]
\item Success Ratio : It is the ratio of number of successful payment to the total number of payment transfer request submitted in an epoch. 
\item TTR \textit{(Time Taken to Route)} : Given a payment transfer request, it is the time taken from the start of routing protocol till its completion (returning the set of feasible paths).
\end{itemize}
We allow just one trial (i.e. $a=1$) of \textit{SpeedyMurmur} since \textit{HushRelay} executes just once. The number of landmarks is varied as 1,2,4 and 6.

\begin{itemize}[leftmargin=*]
\item Real Instances - \emph{HushRelay} and \emph{SpeedyMurmur} has been executed on real instances - Ripple Network \cite{malavolta}, Lightning Network \cite{seres2019topological}. The results are tabulated in Table \ref{tab:1}. 
\item Simulated Instances - The capacity of each payment channel is set between 20 to 100 and each transaction value ranges from 10 to 80. For each synthetic graph, we have executed a set of 2000 transactions, with original state of the graph being restored after a transaction gets successfully executed. The source code for SpeedyMurmur is available in \cite{crysp}.  It is written in Java and makes use of the graph analysis tool GTNA\footnote{https://github.com/BenjaminSchiller/GTNA}.  From the graphs plotted in Fig. \ref{fig:example2} a) and b), it is seen that as the number of landmarks increases, SpeedyMurmur gives better success ratio but at the cost of delayed routing. On the other hand, our routing algorithm, which is independent of any landmark, achieves a better success ratio in less time. 
\end{itemize}
From the results, we can infer that random splitting of capacity without any knowledge of residual graph may lead to failure in spite of presence of routes with the required capacity. 

\section{Conclusion}
\label{7}
In this paper, we have proposed a novel privacy preserving routing algorithm for payment channel network, \textit{HushRelay} suitable for simultaneous payment across multiple paths. From the results, it was inferred that our proposed routing algorithm outperforms landmark based routing algorithms in terms of success ratio and the time taken to route. Currently all our implementations assume that the network is static. In future, we would like to extend our work for handling dynamic networks as well. Our algorithms have been defined with respect to a transaction between a single payer and payee but it can extended to handle multiple transaction by enforcing blocking protocol or non blocking protocol to resolve deadlocks in concurrent payments. \cite{malavolta}.

\bibliographystyle{IEEEtranS}
\bibliography{PCN}

\end{document}